\documentclass{article}
\usepackage{amssymb}
\usepackage{amsfonts}
\usepackage{amsmath}
\usepackage{latexsym}

\begin{document}

\title{On scattering of electomagnetic waves by a wormhole }
\author{A.A. Kirillov, E.P. Savelova \\
Dubna International University of Nature, Society and Man,\\
Universitetskaya Str. 19, Dubna, 141980, Russia }
\date{}
\maketitle

\begin{abstract}
We consider scattering of a plane electromagnetic wave by a wormhole. It is
found that the scattered wave is depolarized and has a specific interference
picture depending on parameters of the wormhole and the distance to the
observer. It is proposed that such features can be important in the direct
search of wormholes.
\end{abstract}

1. As it was shown recently all features of cold dark matter models (CDM)
can be reproduced by the presence of a gas of wormholes \cite{KS11}. At very
large scales wormholes behave exactly like very heavy particles, while at
smaller subgalactic scales wormholes strongly interact with baryons and cure
the problem of cusps. Therefore, we may state that up to date wormholes give
the best candidate for dark matter particles. The final choice between
different dark matter candidates seem to require a direct observation of a
cosmological wormhole. Possible observational effects of wormholes attract
the more increasing attention e.g. see Refs. \cite{whtests} and references
therein. In this paper we consider the most simplest effect which may be
used in observations. We demonstrate that the scattering of a plane
electromagnetic wave by a wormhole leads to a partial depolarization and a
specific interference picture in the scattered signal. The interference
picture depends on the distance to the observer and specific parameters of a
wormhole giving thus an instrument for the possible search of wormholes.

2. The simplest wormhole is described by the metric%
\begin{equation}
ds^{2}=c^{2}dt^{2}-h^{2}\left( r\right) \delta _{\alpha \beta }dx^{\alpha
}dx^{\beta },  \label{wmetr}
\end{equation}%
where
\begin{equation}
h\left( r\right) =1+\theta \left( b-r\right) \left( \frac{b^{2}}{r^{2}}%
-1\right)
\end{equation}%
and $\theta \left( x\right) $ is the step function. Such a wormhole has
vanishing throat length. Indeed, in the region $r>b$, $h=1$ and the metric
is flat, while the region $r<b$, with the obvious transformation $y^{\alpha
}=\frac{b^{2}}{r^{2}}x^{\alpha }$, is also flat for $y>b$. Therefore, the
regions $r>b$ and $r<b$ represent two Minkowski spaces glued at the surface
of a sphere $S^{2}$ with the centre at the origin $r=0$ and radius $r=b$.
Such a space can be described with the ordinary double-valued flat metric in
the region $r_{\pm }>b$ by
\begin{equation}
ds^{2}=c^{2}dt_{\pm }^{2}-\delta _{\alpha \beta }dx_{\pm }^{\alpha }dx_{\pm
}^{\beta },  \label{wmetr2}
\end{equation}%
where the coordinates $x_{\pm }^{\alpha }$ describe two different sheets of
space.

A generalization appears when we change the step function $\theta \left(
x\right) $ with any smooth function $\widetilde{\theta }\left( x\right) $
which has the property $\widetilde{\theta }\left( x\right) \rightarrow 0$ as
$x\rightarrow -\infty $ and $\widetilde{\theta }\left( x\right) \rightarrow
1 $ as $x\rightarrow 0$. In confront to the previous case, such a wormhole
will have a non-vanishing throat length. However, while it differs in
details, the general features remain the same. Such wormholes have vanishing
mass, but one may also include a non-vanishing wormhole mass as described by
\cite{Vis}. In particular, as it was shown in Ref. \cite{KS11} the
background density of baryons automatically generates a some finite rest
mass value. Save the exotic matter which is necessary to support the
cosmological wormhole, it (as a massive object) has also to be surrounded
with standard forms of matter (dust, gas, etc.) which may complicate the
direct observation of wormholes.

Identifying the inner and outer regions of the sphere $S^{2}$ allows the
construction of a wormhole which connects regions in the same space (instead
of two independent spaces). This is achieved by gluing the two spaces in (%
\ref{wmetr2}) by motions of the Minkowski space (the Poincare motions). If $%
R_{+}^{\mu }=(t_{+},\vec{R}_{+})$, where $\vec{R}_{+}$ is the position of
the sphere in coordinates $x_{+}^{\mu }$, then the gluing is the rule%
\begin{equation}
x_{+}^{\mu }=R_{+}^{\mu }+\Lambda _{\nu }^{\mu }\left( x_{-}^{\nu
}-R_{-}^{\nu }\right) ,  \label{gl}
\end{equation}%
where $\Lambda _{\nu }^{\mu }\in SO(3,1)$, which represents the composition
of a translation and a Lorentz transformation of the Minkowski space. In
terms of common coordinates such a wormhole represents the standard flat
space in which the two cylinders $S_{\pm }^{2}\times R^{1}$ (with centers at
positions $\mathbf{R}_{\pm }$) are glued by the rule (\ref{gl}). Thus, in
general, the wormhole is described by a set of parameters: the throat radius
$b$, positions and velocities of throats $\mathbf{R}_{\pm }$ , $\mathbf{V}%
_{\pm }$ , spatial rotation matrix $U_{\beta }^{\alpha }\in O(3)$, and in
general some additional time shift $\Delta t=t_{+}-t_{-}$. For the sake of
simplicity we, in the present paper, suppose that throats does not move in
space $\mathbf{V}_{\pm }\approx 0$ and the time shift is absent $\Delta
t\approx 0$. The motion of one throat (say $S_{+}$) can be excluded by the
choice of the reference system, while the possible respective motion of the
second throat is supposed to have a small non-relativistic velocity $\mathbf{%
V}_{-}\ll c$.

3. The problem of the scattering of radiation on a static gas of
wormholes has been considered recently in Ref. \cite{KSS09}. In
particular it was demonstrated that any discrete source turns out
to be surrounded with a diffuse halo which should be correlated
with analogous halo of dark matter. However, we used there an
approximation in which the size of wormholes was negligible. Here
we consider some general features of the scattering by a wormhole
of a finite size and account for the vector character of
radiation. We point out Ref. \cite{sct} where some aspects of the
scattering by an Ellis geometry were also considered.

Scattering of electromagnetic waves in the short-wavelength limit by a
conducting sphere is the classical problem of electrodynamics \cite{jackson}%
. We do not repeat all derivations here referring to the above textbook but
explain additional features which come from a wormhole.

Let $\mathbf{E}=\mathbf{E}_{0}e^{-i\omega t+i\mathbf{k}_{0}\mathbf{r}}$ be
incident field, then the surface of a sphere can be divided into illuminated
and shadow regions which produce contributions to the scattered field as
\begin{equation}
\mathbf{E}_{sh}\approx ikb{^{2}}\frac{J_{1}(kb\sin \theta )}{kb{\sin \theta }%
}\frac{e^{-i\omega t+ikr}}{r}\frac{\left[ \left[ \mathbf{kE}_{0}\right]
\mathbf{k}\right] }{k^{2}}
\end{equation}%
from the shadow region and
\begin{equation}  \label{il}
\mathbf{E}_{ill}\approx -\frac{b}{2}E_{0}\frac{e^{-i\omega t+ikr}}{r}%
e^{-2ikb\sin \frac{\theta }{2}}\mathbf{e}_{0}
\end{equation}%
form the illuminated region respectively. Here $b$ is the radius of the
sphere,
\begin{equation*}
\mathbf{e}_{0}=(2(\mathbf{n}_{0}\mathbf{E}_{0})\mathbf{n}_{0}-%
\mathbf{E}_{0})/E_{0},
\end{equation*}%
$\mathbf{n}_{0}$ is the unit vector along $\mathbf{k}%
-\mathbf{k}_{0}$, $\mathbf{k}$ $=k\mathbf{r}/r$ and $\cos \theta =(\mathbf{kk%
}_{0})/k^{2}$. It is also supposed that the sphere is at the origin. The
contribution from the shadow region represents the standard diffraction
which does not depend on the nature of the obstacle (it depends on the
projected area only) and gives a very narrow beam along the incident signal $%
\theta \lesssim 10/kb\,$, while the illuminated region gives an isotropic
intensity of radiation.

Now we add features from the structure of a wormhole. It turns out that a
general wormhole can be considered as a couple of dielectric spheres $S_{+}$
and $S_{-}$ glued along the surface. Thus the incident signal generates
illuminated and shadow regions on the outer side of surface and analogous
illuminated and shadow regions on the inner side of the surface. Due to the
gluing the inner side of $S_{+}$ corresponds to the outer side of $S_{-}$
and vise versa. Since points on the spheres are glued both throats radiate
the scattered signal in the same manner and we may consider radiation from
one throat. The incident field is partially reflected by the throat (in the
illuminated region) and partially goes through the throat. The reflection
and transmittance coefficients $\left\vert r\right\vert ^{2}+\left\vert
t\right\vert ^{2}=1$ depend on the specific structure of the wormhole and,
in particular, on the matter content filling the throat and surrounding the
wormhole (E.g., in the vacuum case $r(b,k)\rightarrow 0$ as $bk\gg 1$, e.g.,
see Refs. \cite{sct}). Thus the total scattered signal comprises an
additional term as follows
\begin{equation}
\mathbf{E=E}_{sh}+r\mathbf{E}_{ill}+t^{\prime }\mathbf{E}_{ill}^{\prime }.
\end{equation}%
Let $S_{+}$ be the throat at the origin. The additional term has the same
form as (\ref{il}) and describes the wave transmitted through the throat
(i.e., radiation of the wave adsorbed on the conjugated throat $S_{-}$).
Such term is equivalent to the classical reflection by the sphere $S_{+}$ of
an additional wave $\mathbf{E}^{\prime }=\mathbf{E}_{0}^{\prime }e^{-i\omega
^{\prime }t+i\mathbf{k}_{0}^{\prime }\mathbf{r+i}\psi }$ where $\psi =%
\mathbf{-}\omega \Delta t\mathbf{+\mathbf{k}_{0}\Delta R}$ ($\mathbf{\Delta
R=R}_{-}-\mathbf{R}_{+}$) is the phase difference which the incident field
has in the center of the throat $S_{-}$, and $\mathbf{E}_{0}^{\prime }$, $%
\omega ^{\prime }$, and $\mathbf{k}_{0}^{\prime }$ are related to $\mathbf{E}%
_{0}$, $\omega $, and $\mathbf{k}_{0}$ by the Lorentz transformation $%
\Lambda _{\nu }^{\mu }$ which defines the gluing (\ref{gl}).

The shadow contribution $\mathbf{E}_{sh}$ gives a very narrow beam along $%
\mathbf{k}_{0}$, while the illuminated parts define an omnidirectional flux $%
\mathbf{E}_{ill}^{tot}=r\mathbf{E}_{ill}+t^{\prime }\mathbf{E}_{ill}^{\prime
}$ which is depolarized (since in general $\mathbf{e}_{0}\neq \mathbf{e}%
_{0}^{\prime }$). In the case when the wormhole is surrounded with a dense
plasma $t^{\prime }\rightarrow 0$ (as $bk\gg 1$) and the scattered signal
does not differ from the reflection by a conducting sphere \cite{jackson},
while in the vacuum case $r\rightarrow 0$ (as $bk\gg 1$) \cite{sct} and the
scattered signal corresponds to the reflection by a conducting sphere of the
auxiliary wave $\mathbf{E}^{\prime }$ which relates to the incident wave by
the Lorentz transformation (\ref{gl}). In the last case the scattered signal
has, in general, a somewhat different wavelength $k^{\prime }$ and a
different polarization $\mathbf{e}_{0}^{\prime }$ as compared to the
standard conducting sphere. In other words, we may state that in comparision
to the scattering by a conducting sphere the wormhole leads to some
additional shift of the wavelength and a rotation of the polarization. In
the most general intermediate case the scattered signal shows a more complex
interference picture. The intensity of the flux $I=\left( \mathbf{EE}^{\ast
}\right) $ is given by%
\begin{equation}
I=\frac{b^{2}}{4}\frac{E_{0}^{2}}{r^{2}}\left( 1+A\cos \left\{ \psi -\Delta
k\left( ct-r\right) -\left( 2kb\sin \frac{\theta }{2}-2k^{\prime }b\sin
\frac{\theta ^{\prime }}{2}\right) \right\} \right) .  \label{int}
\end{equation}%
Here $\cos \theta =\left( \mathbf{rk}_{0}\right) /rk_{0}$, $\cos \theta
^{^{\prime }}=\left( \mathbf{rk}_{0}^{\prime }\right) /rk_{0}$, $\Delta
k=c\left( \omega -\omega ^{\prime }\right) $, and $A=(rt^{\prime \ast
}+r^{\ast }t^{\prime })\left( \mathbf{e}_{0}\mathbf{e}_{0}^{\prime }\right) $

We see that the most strong interference picture appears when $A\simeq 1$
(recall that $A$ depends on the specific structure and the matter
composition of the wormhole) and is given by the phase $\Delta k\left(
ct-r\right) $ which comes from the possible respective motions of throats $%
\Delta \omega \simeq k\Delta V$ (here $\mathbf{\Delta V=V}_{+}-\mathbf{V}_{-}
$ and $c$ is the speed of light) and $\Delta \omega $ is he standard Doppler
shift. In this case the phase $\psi $ has also a dependence on time (via $%
\mathbf{R}_{\pm }\left( t\right) $). However the additional phase in (\ref%
{int}) forms a more peculiar and complex interference picture with a much
larger wavelength $\delta k\ll \Delta k$ which in principle can be measured
while the Earth orbiting the Sun. We point out that these oscillations are
not pure harmonic ones.

Consider for simplicity the case when the Lorentz transformation reduces to
a pure spatial rotation $U_{\beta }^{\alpha }$ and the frequency remains the
same $\omega ^{\prime }=\omega $. Then expanding (\ref{int}) by $\delta
\mathbf{r}$ near the observer at the position $\mathbf{r}$ we find the
expression
\begin{equation*}
I=\frac{b^{2}}{4}\frac{E_{0}^{2}}{r^{2}}\left( 1+A\cos \left\{ \varphi
_{0}+\left( \delta \mathbf{k}\delta \mathbf{r}\right) \right\} \right)
\end{equation*}%
where the constant phase $\varphi _{0}=\psi -2kb\left( \sin \frac{\theta }{2}%
-\sin \frac{\theta ^{\prime }}{2}\right) $, and
\begin{equation}
\delta \mathbf{k}=\left[ \frac{1}{sin\frac{\theta }{2}}-\frac{1}{sin\frac{%
\theta ^{\prime }{}}{2}}\right] \frac{2b}{r}\mathbf{k}_{0}.  \label{dk}
\end{equation}%
Thus the intensity has a specific additional oscillations with a very big
wavelength proportional to the ratio $\lambda _{0}r/b\gg \lambda _{0}$ where
$\lambda _{0}$ is the wavelength of the incident signal. We suppose that the
interference picture described above should be taken into account in
analyzing observations and may be useful in the search of wormholes.

This research was supported in some part by RFBR 09-02-00237-a and by a
Royal Society grant

\end{document}